\documentstyle[12pt]{article}
\def\np    { Nucl. Phys. }

\def\pl    { Phys. Lett. }

\def\begineq{\begin{equation}}
\def\endeq{\end{equation}}

\def\beqa{\begin{eqnarray}}
\def\eeqa{\end{eqnarray}}

\def\parn              {  \par\noindent }
\def\parbigskip        {  \par\bigskip  }

\def\parbigskipn        {  \par\bigskip\noindent  }

\def\parsmallskipn      {  \par\smallskip\noindent  }
\def\papertitlepage{\baselineskip 3.5ex \thispagestyle{empty}}
\def\Title#1{\vspace{2.5cm}\begin{center}
 {\LARGE\bf #1} \end{center} 
\vspace{2cm}}
\def\Authors#1{\begin{center} {\Large\it #1} \end{center}}
\def\Abstract{\vspace{2.0cm}\begin{center} {\Large\bf Abstract} 
           \end{center} \parbigskip}
\def\ICRRnumber#1#2#3{\hfill \begin{minipage}{4cm} ICRR-#1
              \parn #2 \parn #3 \end{minipage}}

\begin{document}
\papertitlepage
\vspace*{-1 cm}
\ICRRnumber{416-98-12}{April 1998}{hep-th/9804202}
\Title{The Spacetime Superalgebras   \\ 
\vskip 1.5ex from M-branes \\
\vskip 1.5ex in M-brane Backgrounds} 
\Authors{{\sc\  Takeshi Sato
\footnote{tsato@icrr.u-tokyo.ac.jp}} \\
 \vskip 3ex
 Institute for Cosmic Ray Research, University of Tokyo, \\
 Tanashi, Tokyo 188-8502 Japan \\
  }
\Abstract
We derive the spacetime superalgebras 
explicitly from ``test'' M-brane actions 
in M-brane backgrounds 
to the lowest order in $\theta$ via canonical formalism,
and discuss various BPS saturated configurations
on the basis of their central charges
which depend on the harmonic functions determined by
the backgrounds.
All the 1/4 supersymmetric intersections of two M-branes
obtained previously are deduced
from the requirement  
of the test branes to be so ``gauge fixed''
in the brane backgrounds as to preserve 1/4 supersymmetry.
Furthermore, some of 1/2-supersymmetric bound states
of two M-branes
are deduced from the behavior of the harmonic functions
in the limits of vanishing distances of the two branes.
The possibilities of some triple intersections
preserving 1/4 supersymmetry are also discussed.
          
\newpage

%

\baselineskip=0.6cm


\section{Introduction}

%

The M-theory is currently a most hopeful candidate for a unified
theory of particle interactions\cite{tow2}\cite{wit1}
and is extensively studied from various points of
view\cite{tow3}\cite{sus1}.
Among them,
the analysis via superalgebra is one of the most powerful approaches
to investigate its various properties\cite{hul1}\cite{tow7}. 
Since there are, of cource, two kinds of supersymmetries,
two kinds of algebras have been discussed so far:
the spacetime superalgebra and the worldvolume ones.
The former is initially constructed as the most general modification
of the standard D=11 supersymmetry algebra\cite{hol1}\cite{tow4},
and then 
deduced explicitly from the M-5-brane action in the {\it flat}
background via (anti-)canonical commutation relations of the
worldvolume fields\cite{tow5}(see also ref.\cite{asc1}).
It is called M-theory superalgebra,
from which it is shown that only five basic
1/2 supersymmetric constituents of M-theory can be permitted.
And the latters, defined on the flat (p+1)-dimensional
worldvolumes of p-branes, are the 
maximal extentions of the (p+1)-dimensional
supertranslation algebras,
from which all the 1/4 supersymmetric M-brane intersections 
as worldvolume solitons can
be deduced \cite{how1}\cite{tow6}\cite{tow10}.
Both of these analyses are also applied to
D-branes\cite{tow6}\cite{ham1},
although there are some
subtleties in the worldvolume cases.
In this way these discussions so far have been based
only on the flat spacetime.

Now, we will discuss the extension, in particular, of the former
to those from branes in {\it nontrivial}
backgrounds.\footnote{
The possibility of this extension has been already
pointed out 
in the earlier paper\cite{asc1}
for a different purpose(related to nontrivial topologies),
although it is not shown explicitly there.}
One of our most important motivation for it is 
to get the superalgebra of
the 10-dimensional
massive IIA theory\cite{berg6}\cite{berg7}\cite{berg5}.
It has not been obtained yet 
since this theory does not admit the flat background
because of the existence of
cosmological constant\cite{rom1}\cite{berg3}.
This superalgebra is significant 
not only to investigate the properties of the theory
like the above cases,
but also
to understand the 11-dimensional origin of
the massive IIA theory, 
which is not yet known satisfactorily
because of the no-go theorem presented in ref.\cite{des1},
although several trials were
made\cite{how4}\cite{berg4}\cite{loz1}(and recently \cite{berg9}).  
Therefore, the extension is urgently necessary.

Let's reconsider the computation in ref.\cite{tow5}. 
We might be able to interpret it as follows:
the M-5-brane action is originally invariant
under {\it local} super-transformation 
before taking its background to be any specific geometry.
Suppose we float the M-5-brane as a ``test'' brane 
in the flat background.
Then, the system and hence the action have
the supertranslation symmetry in all the directions 
for the following two reasons:
first, the background has the supertranslation symmetry and 
the test brane is assumed to be so light that 
it has no effect on the background.
Second, the configuration of the test brane 
including its orientation 
has not yet been fixed at this time. 
Therefore, we can define the Noether supercharge, compute its
anti-commutator as the superalgebra and discuss
the supersymmetric configurations
permitted in the flat background on the basis of
its central charges,
which was done in ref.\cite{tow5}. 

This interpretation will suggest us to think of 
the following extention:
if we float a test M-brane in a certain {\it non-trivial} background 
which has some portions of supersymmetry,
then the system and the action {\it should} have the same supersymmetry
\footnote{In the middle 
of completing this work, this idea is
pointed out and proved generically by ref.\cite{berg1}
in some other context
(about the new actions presented in it). 
Our work 
will be worth doing in order to examine this idea explicitly
to low orders in $\theta$.}
because of the same reasons stated above.
Therefore, in the same way, we can define the corresponding
Noether supercharge,
compute its commutator as the superalgebra,
and finally, we are
{\it able to deduce} from it
{\it the supersymmetric configurations of the test brane 
(including intesections with the background) 
permitted in the non-trivial background}.

The aim of this paper is to examine this idea explicitly
in the cases of a test M-2-brane and a test M-5-brane
in an M-2-brane and an M-5-brane background,
i.e. four cases in all.\footnote{We apply the method to the case in
10D massive IIA background in the next paper\cite{sato1},
which will be completed soon.}
The concrete procedures are as follows: 
we will take the background to be a M-brane solution
which actually consists of
such a large number of coincident M-5-branes
that our ``test brane'' approximation is justified,
and substitute the solution for the test brane action as was done
in ref.\cite{tow8}.
Then, we
check the invariance of the action under the unbroken supersymmetry
transformation, derive the representation of the supercharge 
in terms of the worldvolume fields of the
test brane and their conjugate momenta,
compute its algebra
and discuss various supersymmetric configurations
permitted in the background on the basis of
the algebra.

We note that our computations are performed only up to the low
orders in $\theta$ which might contribute to the central charges 
at zeroth order in $\theta$
because they suffice to discuss the supersymmetric configurations
which we want to know.
The most important fact throughout our computations is that
we can reduce the superspace with the supercoordinates $(x,\theta)$ 
to that with coordinates $(x,\theta^{+})$,
where the sign $+$ of $\theta^{+}$ implies
that $\theta^{+}$ has a definite
worldvolume chirality of the background. 
The reason is the following: 
since half of supersymmetry is already not the symmetry of the system 
owing to the existence of the background brane,
the corresponding parameter $\theta^{-} $ must not be transformed.
Thus, the conjugate momentum of $\theta^{-} $ does not appear in the
supercharge $Q^{+}$, which means that the terms
including $\theta^{-} $ do not contribute to the
central charges at zeroth order in $\theta$.
Therefore, {\it we set $\theta^{-}=0 $ from the biginning}.

The consequence is that
all the 1/4-supersymmetric intersections of two M-branes
obtained previously both in 11D supergravity\cite{tow8}\cite{gau1} and 
via worldvolume superalgebras\cite{tow9} are {\it deduced}
from the requirement  
of the test branes to be so ``gauge fixed'' in the backgrounds
as to preserve 1/4-supersymmetry.
In addition, one outstanding characteristic of the results is
the dependence of (the r.h.s. of) the superalgebras
on the harmonic functions determined by the backgrounds.
By using this property
we derive the following two kinds of bound states
composed of two M-branes preserving 1/2 supersymmetry:
the first ones are deduced from the configurations of
the test M-p-branes parallel to the background M-p-brane
with the converse orientation. The second is from
the configuration of the test M-2-branes parallel to 
a two-dimensional subspace of the worldvolume
of the background M-5-brane. 
In both cases all the supersymmetry is broken because
there are only attractive forces 
between the two branes.
Then, the ``absorption'' limits, namely the 
limits of zero distances are expected to lead to the restorations
of 1/2 spacetime supersymmetry
because the potential energies are minimized in the limit.
It is shown that these are really the cases,   
which correspond to M-p-brane/M-p-brane
bound states and
a M-2-brane/M-5-brane bound state\cite{izq1}\cite{tow5}
(preserving 1/2 supersymmetry), respectively.
Another merit of the results is that
the possibilities of some supersymmetric triple
intersections can be discussed directly
on the basis of the spacetime superalgebras,
although not systematically.

This paper is organized as follows:
in section 2 we discuss the superalgebras from the test M-2-brane
in an M-2-brane and an M-5-brane background.
In section 3 we discuss the superalgebras from the test M-5-brane
in an M-2-brane and an M-5-brane background.
In section 4 we give short summary and discussions.  

Before presenting our results we give some preliminaries.
We use ``mostly plus'' metrics for both worldvolume and spacetime.
And we use Majorana ($32 \times 32$) representation for Gamma matrices 
$\Gamma_{\hat{m}}$ which are all real and satisfy 
$ \{ \Gamma_{\hat{m}} , \Gamma_{\hat{n}} \} 
= 2\eta_{\hat{m}\hat{n}}$. 
$\Gamma_{\hat{0}}$ is antisymmetric and others 
symmetric. Charge Conjugation is ${\cal C}=\Gamma_{\hat{0}}$.
We use the symbol $\natural$ to denote the number 10 as used in 
\cite{tow7} .
We use capital latin letters($M,N,..$) for superspace indices, 
small latin letters($m,n,..$) for spacetime vectors 
and early small greek letters
($\alpha,\beta$,..) for spinors. Furthermore, we use late greek letters     
($\mu,\nu,..$) for spacetime vectors paralell to the background branes and
early latin letters($a,b,..$) for spacetime vectors transverse to them.
We use {\it hatted letters} ($\hat{M},\hat{m},\hat{a},\hat{\alpha}..$) 
for {\it all the inertial frame indices}
and finally middle latin letters($i,j,..$) for worldvolume vectors.

%

\section{The Superalgebras from the M-2-brane in M-brane Backgrounds }

In this section we will deal with the M-2-brane in M-brane
backgrounds.  

(2a) the M-2-brane in an M-2-brane background

At first we will begin with the case of the test M-2-brane floating
in an M-2-brane background.
The M-2-brane action in a D=11 supergravity background is\cite{berg2} 
\beqa
S_{M2}=-T\int d\xi^{3}\sqrt{-{\rm det}g_{ij}}
+T\int d\xi^{3} \frac{1}{3!}\varepsilon^{ijk}C^{(3)}_{ijk}
\label{M2ea} 
\eeqa
where $g_{ij}=E_{i}^{\hat{m}}E_{j}^{\hat{n}}\eta_{\hat{m}\hat{n}}$
is the induced worldvolume metric and $C^{(3)}_{ijk}$ is 
the worldvolume 3-form induced by the superspace 3-form gauge
potential.
$E_{i}^{A}=\partial_{i}Z^{M}E_{M}^{\ \ \hat{A}}$
where $E_{M}^{\ \ \hat{A}}$ is the supervielbein. 
Note that at this time the action is invariant under local 
super-transformation.

Let's fix the background to an M-2-brane solution given by\cite{duf1}
\beqa
ds_{11}^{2} &=& H^{-2/3}\eta_{\mu\nu}dx^{\mu}dx^{\nu}
+H^{1/3}dy^{a}dy^{b}\delta_{ab}
\nonumber\\
C_{\mu_{1}\mu_{2}\mu_{3}} &=&
\frac{\varepsilon_{\mu_{1}\mu_{2}\mu_{3}}}{{\rm det}g_{\mu\nu}}H^{-1},
\ \ \ \ {\rm (the \ others) }=0 \label{M2back}
\eeqa
where $\eta_{\mu\nu}$ is the 3-dimensional Minkovski metric with
coordinates $x^{\mu}$ and H is a harmonic function 
on the transverse 8-space with coordinates $y^{a}$, that is,
$H=1+\frac{q_{2}}{y^{6}}$ where $y=\sqrt{y^{a}y^{b}\delta_{ab}}$
and $q_{2}$ is a constant. 
$\varepsilon_{\mu_{1}\mu_{2}\mu_{3}}
=g_{\mu_{1}\nu_{1}}g_{\mu_{2}\nu_{2}}g_{\mu_{3}\nu_{3}}
\varepsilon^{\nu_{1}\nu_{2}\nu_{3}}$ and $\varepsilon^{012}=1$.
This background admits a Killing spinor $\varepsilon$
which satisfies 
\beqa
\delta\psi_{m}
=(\partial_{m}+\frac{1}{4}\omega_{m}^{\ \hat{r}\hat{s}}
\Gamma_{\hat{r}\hat{s}}
+T_{m}^{\ n_{1}n_{2}n_{3}n_{4}}
F_{n_{1}n_{2}n_{3}n_{4}})\varepsilon =0
\eeqa
where $T_{m}^{\ n_{1}n_{2}n_{3}n_{4}}
=-\frac{1}{288}(\Gamma_{m}^{\ n_{1}n_{2}n_{3}n_{4}}
+8\Gamma^{ [ n_{1}n_{2}n_{3}}\delta_{m}^{n_{4} ] })$.
Then the Killing spinor has the form 
$ \varepsilon =H^{-1/6}\varepsilon_{0}$
where $\varepsilon_{0}$ has the  positive worldvolume chirality, i.e.
$\bar{\Gamma}\varepsilon_{0}\equiv \Gamma_{\hat{0}\hat{1}\hat{2}}
\varepsilon_{0}=+\varepsilon_{0}$.

Since $\bar{\Gamma}$ satisfies 
$\bar{\Gamma}^{T}=\bar{\Gamma}$ and $\bar{\Gamma}^{2}=1$,
both $\frac{1\pm\bar{\Gamma}}{2}$ and
$\frac{1\pm\bar{\Gamma}^{T}}{2}$ are projection operaters.
So, if we denote 
$\ \frac{1\pm\bar{\Gamma}}{2}\zeta$ as $\zeta^{\pm}$
for a spinor $\zeta $,
the background 
is invariant under the transformation 
generated by the supercharge $Q^{+}$ 
and so is the system because the negligibly light
test brane
is assumed not to affect the background geometry 
and its configuration is not fixed yet at this moment.
On the other hand, the background and the brane action 
are not invariant under the transformation by $Q^{-}$,
which means that we should set the corresponding transformation parameter
$\varepsilon^{-}$ to be zero.
Then, the conjugate momentum $\Pi^{-}$ 
of $\theta^{-}$ does not appear in the Noether charge $Q^{+}$ 
only whose algebra we are interested in.
Therefore, the terms including $\theta^{-}$ {\it never} contribute to 
the central charges at zeroth order in $\theta$.
Thus, we can set from the beginning
\beqa
\theta^{-}=0. \label{thetam}
\eeqa
From now on, we will use these freely in all the cases we treat 
in this paper. Related with this,
we exhibit the properties of $\bar{\Gamma}$: 
\beqa
[ \bar{\Gamma},\Gamma_{\hat{\mu}} ] = [ \bar{\Gamma},{\cal C} ]
= \{ \bar{\Gamma},\Gamma_{\hat{a}}\} = 0. \label{gammabar}
\eeqa

Now, we are prepared to get the explicit representations of
the superfields and the super-coordinate transformation
in terms of component fields
to low orders in $\theta$.
By substituting (\ref{M2back}) to the usual expressions\cite{cre1}
and using (\ref{thetam}) and (\ref{gammabar})
we see that only the $E_{a}^{\ \hat{\alpha}}$ has 
the nontrivial contribution from the background.
From the results
the superspace 1-form on the inertial frame
$E^{\hat{A}}=dZ^{M}E_{M}^{\ \ \hat{A}}$ is given by
\footnote{In fact we need to know the (vanishing of the)
contribution from $E_{m}^{\ \hat{n}}$  
at order $\theta^{2}$. 
We can infer its vanishing 
in this specific simple background,    
but the expression of 
$E_{m}^{\ \hat{n}}$ at order $\theta^{2}$ in general background
was obtained
recently\cite{dew1}, 
by which our inference is confirmed.} 
\beqa
E^{\hat{\mu}} &=& dx^{\nu}H^{-1/3}\delta_{\nu}^{\ \hat{\mu}} 
-i\bar{\theta}^{+}\Gamma^{\hat{\mu}}d\theta^{+} +{\cal O}(\theta^{4})
\nonumber  \\
E^{\hat{a}} &=& dy^{b}H^{1/6}\delta_{b}^{\ \hat{a}}+{\cal
O}(\theta^{4})\nonumber  \\
E^{\hat{\alpha}} &=& d\theta^{\hat{\alpha}+}+\frac{1}{6}H^{-1}dH
\theta^{\hat{\alpha}+}+{\cal O}(\theta^{3})
\label{E1M2}.
\eeqa
Since the 1-form $E^{\hat{A}}$ has
no superspace (curved) indices,
$E^{\hat{A}}$, and hence the Nambu-Goto action,
are invariant 
under the super-coordinate transformation\cite{cre1} 
$\delta Z^{M}=\Xi^{M}$ in this background given by
\beqa
\Xi^{\mu} &=&i\bar{\varepsilon}^{+}\Gamma^{\mu}\theta^{+} 
+ {\cal O}(\theta^{3}) \nonumber \\
\Xi^{a} &=& 0+ {\cal O}(\theta^{3})  \nonumber \\
\Xi^{\alpha} &=& \varepsilon^{\alpha +}+ {\cal O}(\theta^{2}).
\label{supertr}
\eeqa
We can easily check the invariance of $E^{\hat{A}}$ explicitly
up to second order in $\theta$.
Note that the coordinates $y^{a}$ transverse to the background brane
are not transformed (at least up to the second order in $\theta $),
i.e. this is the supertranslation paralell to 
the background brane. (Thus, we can define
the corresponding Noether supercharge.)
And it is also to be noted that
$\Gamma^{\mu}=H^{1/3}\Gamma^{\hat{\nu}}
\delta_{\hat{\nu}}^{\mu}$, i.e. the gamma matrices
with the spacetime
indices depend on the harmonic function.

The remaining field is the super-3-form gauge potential.
It is introduced by the gauge invariant 4-form field
strength\cite{cre1}\cite{bri1}
\beqa
R^{(4)}\equiv dC^{(3)}=\frac{i}{2}E^{\hat{m}}E^{\hat{n}}
\bar{E^{\hat{\alpha}}}
(\Gamma_{\hat{n}\hat{m}})_{\hat{\alpha}\hat{\beta}}E^{\hat{\beta}}
+\frac{1}{4!}E^{\hat{m_{1}}}E^{\hat{m_{2}}}E^{\hat{m_{3}}}E^{\hat{m_{4}}}
F_{\hat{m_{4}}\hat{m_{3}}\hat{m_{2}}\hat{m_{1}}} \label{R4M2},
\eeqa
where $F_{\hat{m_{4}}\hat{m_{3}}\hat{m_{2}}\hat{m_{1}}}$ is the bosonic
field strength which is in this case 
associated with the electric M-2-brane background.
From (\ref{R4M2}) we get
\footnote{Although the $\hat{\alpha}$ of $\theta^{\hat{\alpha}+}$
is the index of the inertial frame, 
$\theta^{\hat{\alpha}}
=\theta^{\beta}\delta_{\beta}^{\hat{\alpha}}+{\cal O}(\theta^{3}) $.
So, we need not distinguish the two indices in this paper.} 
\beqa
C^{(3)} &=& \frac{1}{3!}H^{-1}(-\epsilon_{\mu\nu\rho})
dx^{\mu}dx^{\nu}dx^{\rho}
-\frac{i}{2}H^{-2/3}dx^{\rho}\delta_{\rho}^{\hat{\mu}}
dx^{\sigma}\delta_{\sigma}^{\hat{\nu}}\bar
{\theta^{+}}\Gamma_{\hat{\mu}\hat{\nu}}d\theta^{+} \nonumber \\
& &
-\frac{i}{2}H^{1/3}dx^{c}\delta_{c}^{\hat{a}}
dx^{d}\delta_{d}^{\hat{b}}\bar
{\theta}^{+}\Gamma_{\hat{a}\hat{b}}d\theta^{+} 
\ \ + {\cal O}(\theta^{4}) \ \ \ \ (\epsilon_{012}=-1),
\label{C3M2}
\eeqa
and hence the supertransformation of $C^{(3)}$:\footnote{
Throughout this paper we make an assumption that
all the fermionic (but not bosonic) cocycles
in the superspaces are trivial.
Then, the invariance of $R^{(4)}$ under the super-transformation
means that $\delta C^{(3)}$ in {\it any} of the supersymmetric 
backgrounds under the assumption
can be written as certain d-exact forms to {\it full} 
order in $\theta$.}
\beqa
\delta C^{(3)} &=& d(-\frac{i}{2}H^{1/3}dy^{c}\delta_{c}^{\ \hat{a}}
dy^{d}\delta_{d}^{\ \hat{b}}\bar{\varepsilon}^{+}\Gamma_{\hat{a}\hat{b}}
\theta^{+}+ {\cal O}(\theta^{3})) 
\equiv  d(\bar{\varepsilon}^{+}\Delta_{2}) \label{delC3}.
\eeqa

Thus, the M-2-brane action (\ref{M2ea}) is invariant under
(\ref{supertr}) up to total derivative,
and we define 
the Noether supercharge $Q_{\alpha}^{+} $ in the Hamiltonian formulation 
as an integral over the test brane at fixed time
${\cal M}_{2}$, given by\cite{asc1}
\beqa
Q_{\alpha}^{+} &\equiv& Q_{\alpha}^{+(0)}-i\int_{{\cal M}_{2}} 
({\cal C}\Delta_{2})_{\alpha} \nonumber \\ 
&=&\int_{{\cal M}_{2}} d^{2}\xi(i\Pi_{\alpha}^{+}
-\Pi_{\mu}({\cal C}\Gamma^{\mu}\theta^{+})_{\alpha})
-\frac{1}{2}T\int_{{\cal M}_{2}}
 dy^{a}dy^{b}({\cal C}\Gamma_{ab}\theta^{+})_{\alpha}
+{\cal O}(\theta^{3})
\eeqa
where $\Pi_{\mu}$ and $\Pi_{\alpha}^{+}$ are the conjugate momemta of 
$x^{\mu}$ and $\theta^{+}$, respectively, and $Q_{\alpha}^{+(0)}$ is 
the contribution from
the Nambu-Goto action
whose form is almost common to all the p-brane.
In this way we get the superalgebra of $Q_{\alpha}^{+}$ :
\beqa
\{ Q_{\alpha}^{+}, Q_{\beta}^{+}\}= 2\int_{{\cal M}_{2}}
d^{2}\xi\Pi_{\mu}
({\cal C}\Gamma^{\mu})_{\alpha\beta}
+2\cdot\frac{1}{2}T\int_{{\cal M}_{2}}  dy^{a}dy^{b}
({\cal C}\Gamma_{ab})_{\alpha\beta}
+{\cal O}(\theta^{2}).\label{spalg22}
\eeqa
Before discussing this result,
we give the explicit expression of $\Pi_{\mu}$:
\beqa
\Pi_{\mu}&=&\frac{\delta {\cal L}^{(0)}}{\delta \dot{x}^{\mu}}
+\frac{T}{2} \varepsilon^{0ij}
\partial_{i}x^{\nu}\partial_{j}x^{\rho}C_{\mu\nu\rho}^{(3)}
+{\cal O}(\theta^{2}) 
\equiv \Pi_{\mu}^{(0)}+\Pi_{\mu}^{WZ}
\eeqa
where ${\cal L}^{(0)} $ is the Nambu-Goto Lagrangian.

We now discuss the implications of this algebra.
First, if the test brane is oriented paralell to the background brane,
the term like a central charge arises from the $\Pi_{\mu}^{WZ}$,
although the original central charge vanishes.
If we choose the static gauge 
$\partial_{0}x^{\mu}=\delta_{0}^{\ \mu}$ and 
$\partial_{i}x^{0}=\delta_{i}^{\ 0} $,
$\Pi_{\mu}^{(0)}$ and $\Pi_{\mu}^{WZ}$ are obtained respectively as
\beqa
\int_{{\cal M}_{2}} d^{2}\xi \Pi_{\mu}^{(0)} &=& 
T|\int_{{\cal M}_{2}} dx^{1}dx^{2}|H^{-1}  
\delta_{\mu}^{\ 0}+{\cal O}(\theta^{2})\label{pi0} \\
\int_{{\cal M}_{2}} d^{2}\xi \Pi_{\mu}^{WZ}&=&
T\int_{{\cal M}_{2}} dx^{1}dx^{2}H^{-1}
\delta_{\mu}^{\ 0}+{\cal O}(\theta^{2})\label{piwz}.
\eeqa
The symbol of the absolute value in $\Pi_{0}^{(0)}$ is   
due to the Jacobian
originated from
the determinant in ${\cal L}^{(0)} $.  
Thus, we conclude that the parallel configuration with 
a certain orientation
of the test brane has the 1/2 spacetime supersymmetry
and the one with the other orientation breaks 
all the supersymmetry.   
Note that (\ref{pi0}) and (\ref{piwz})
are invariant under the 12-plane rotation and hence the discussion
above holds, as it should be. 
Furthermore, even in the case that all the supersymmetry is broken,
1/2 supersymmetry is restored
in the limit $y\to 0$ (i.e. $H \to \infty$). 
The reason is as follows:
since $\Pi_{0} \propto H^{-1}$ and $\Gamma^{\mu} \propto H^{1/3} $,
the r.h.s. of the superalgebra (\ref{spalg22}) is proportional to
$H^{-2/3}$, hence vanishes in the limit.
In fact this restoration is reasonable 
because both of the forces via graviton and anti-symmetic tensor 
are attractive in this case and
the potential energy is formally minimized 
in this ``absorption'' limit.
This is the M-2-brane/M-2-brane
bound state preserving 1/2 supersymmetry.

On the other hand,
if the test brane is oriented orthogonally to the background brane,
the central charge {\it does} have the nonzero value.
In the static gauge with the test brane
to be fixed, for example, to 34-plane,
the algebra becomes
\beqa
\{ Q_{\alpha}^{+}, Q_{\beta}^{+}\}=2T\int_{{\cal M}_{2}} 
d^{2}\xi H^{1/3}(1-\Gamma_{\hat{3}\hat{4}})_{\alpha\beta},
\eeqa
which means that 1/4 spacetime 
supersymmetry is preserved in this configuration.

We note that the r.h.s. of the superalgebra does not vanish completely
in {\it any} limit if the test brane has at least one
coordinate transverse to the background brane. 
This fact is common to all the other cases. 
The reason is the following:
suppose that
one of the worldvolume coordinates of the test brane $\xi^{1}$
equals to the transverse coordinate $y^{\natural}$.
Then, H is expressed as
$H=1+\frac{q}{((\xi^{1})^{2}+(y^{a'})^{2})^{3}}$ where $a'=3,..,9$.
Thus, $\int d^{2} \xi H^{K} $ does not vanish for any constant $K$,
even in the limit of  $ (y^{a'})^{2} \to 0 $. In other words,
the r.h.s. of the superalgebras do not vanish completely
for the cases of the string intersection and the orthogonal orientation.

Thus, we have derived the superalgebra from
the M-2-brane action in an M-2-brane background.
All the 1/4-supersymmetric intersections and
a 1/2-supersymmetric bound state of two M-2-branes
known before\cite{tow9}\cite{tey1}
have been deduced from the algebra.

(2b)the M-2-brane in an M-5-brane background

Next, we will consider the M-2-brane in an M-5-brane background.
The M-5-brane background solution is given by \cite{guv1}
\beqa
ds_{11}^{2} &=& H^{-1/3}\eta_{\mu\nu}dx^{\mu}dx^{\nu}
+H^{2/3}dy^{a}dy^{b}\delta_{ab}
\nonumber\\
F_{abcd} &=& -\varepsilon_{abcde}\partial_{e} H \label{m5back}
\eeqa
where $\mu =0,1,..,5$ and $a=6,..,9,\natural $.
The Killing spinor $\varepsilon$ has the form
$\varepsilon =H^{-1/12}\varepsilon_{0}$
where $\varepsilon_{0}$ has again the positive chirality of
the worldvolume of the background:
$\bar{\Gamma}'\varepsilon_{0}\equiv 
\Gamma_{\hat{0}\hat{1}\hat{2}\hat{3}\hat{4}\hat{5}}
\varepsilon_{0}=+\varepsilon_{0}$. 
Since $\frac{1\pm\bar{\Gamma}^{'(T)}}{2}$ are again projection operators,
only the supersymmetry corresponding to $Q^{+}$ is the symmetry of 
the background.
Thus, for the same reason 
stated in the case of the M-2-brane background,
only $Q^{+}$ is the symmetry of 
the system and   
we set $\varepsilon^{-}=0$ and hence $\theta^{-}=0 $.
We note that $\bar{\Gamma}'$ satisfies the (anti-)commutators
$ \{ \bar{\Gamma}', {\cal C} \} = 
\{ \bar{\Gamma}',\Gamma_{\hat{\mu}} \}=
 [ \bar{\Gamma}', \Gamma_{\hat{a}} ] = 0.$
By using this relations and the formula presented in ref.\cite{cre1}
the superspace 1-form on the inertial frame 
is given by
\beqa
E^{\hat{\mu}} &=& dx^{\nu}H^{-1/6}\delta_{\nu}^{\ \hat{\mu}} 
-i\bar{\theta}^{+}\Gamma^{\hat{\mu}}d\theta^{+} 
+{\cal O}(\theta^{4})\nonumber \\
E^{\hat{a}} &=& dy^{b}H^{1/3}\delta_{b}^{\ \hat{a}}+{\cal
O}(\theta^{4}) \nonumber \\
E^{\hat{\alpha}} &=& d\theta^{\hat{\alpha}+}+\frac{1}{12}H^{-1}dH
\theta^{\hat{\alpha}+}+{\cal O}(\theta^{3}).
\eeqa
The super-coordinate transformation is formally the same form as 
that in the M-2-brane background (\ref{supertr}) except for 
the ranges of $\mu$ and $a$. 
The superspace 3-form $C^{(3)}$ is introduced in the same way 
as (\ref{R4M2}).
Note that $C^{(3)}$ cannot be expressed globally in this
case because 
it is associated with the magnetic M-5-brane solution.
However, neither does it contribute to $\delta {\cal L}^{WZ}$
nor the $\Pi_{\mu}$ up to first
order in $\theta$, owing to the inertness of the
transverse coordinates $y^{a}$ under the super-transformation.
As a result,
\beqa
\delta C^{(3)}=d(-iH^{1/6}dx^{\nu}\delta_{\nu}^{\hat{\mu}}
dy^{b}\delta_{b}^{\hat{a}}\bar{\varepsilon}^{+}
\Gamma_{\hat{\mu}\hat{a}}\theta^{+}+{\cal O}(\theta^{3}))
\equiv d(\bar{\varepsilon}^{+}\Delta_{2}').
\eeqa
Thus, the action is invariant under the super-transformation up to
total derivative.
After the same procedures as in the M-2-brane background, 
we get the expression of the 
supercharge $Q_{\alpha}^{ +} \equiv Q_{\alpha}^{+(0)}-i\int_{{\cal M}_{2}} 
({\cal C}\Delta_{2}')_{\alpha}$.
Then, the superalgebra is obtained as
\beqa
\{ Q_{\alpha}^{+}, Q_{\beta}^{+}\}= 2\int_{{\cal M}_{2}}
d^{2}\xi\Pi_{\mu}
({\cal C}\Gamma^{\mu})_{\alpha\beta}
+2\cdot T\int_{{\cal M}_{2}}dx^{\mu}dy^{a}
({\cal C}\Gamma_{\mu a})_{\alpha\beta}+{\cal O}(\theta^{2})
.\label{spalgmsm5}
\eeqa

We conclude from the form of its
central charge that the string intersection of the test brane
with the background is the only 1/4-supersymmetric
configuration permitted in this background. 
This is consistent with ref.\cite{tow9}\cite{tey1}, too. 
In this case the interpretation of the ``boundary''
of the test M-2-brane is as follows, as given in ref.\cite{tow11}:
if we choose the gauge $\xi^{1}=x^{1}$ and $\xi^{2}=y^{\natural} $,
the intersection on the
hypersurface $y\equiv \sqrt{y^{a}y^{b}\delta_{ab}}=0$, (i.e.,
$\xi^{2}=0$)
does not correspond to any points of the M-2-brane
because the proper distance on it is infinite. 
So, the ``edge'' of the M-2-brane is interpreted to
disappear down the infinite M-5-brane throat.
In other words {\it the test M-brane has no boundary in this method.}
\footnote{If we set
the worldvolume coordinate $\xi^{2}$
to take values in the open interval $(0,\infty)$,
it is interpreted as the M-2-brane ending on the
M-5-brane\cite{tow11},
on so large scales compared to that determined
by M-5-brane tension
that the background solution can be replaced
with the M-brane source.}

On the other hand,
if the test M-2-brane is parallel to any two-dimensional
subspaces of the worldvolume
of the background M-5-brane, all the supersymmetry is broken.
However, 1/2 spacetime supersymmetry is restored
in the limit $y \to 0$ (i.e. $H \to \infty$) since
$\Pi_{0}{\cal C}\Gamma^{0} \propto H^{-1/3}$. 
Thus, we can deduce the M-2-brane/M-5-brane bound state
with 1/2 spacetime supersymmetry given in ref.\cite{izq1}\cite{tow5}.

\section{The Superalgebras from the M-5-brane in M-brane Backgrounds }

In this section we discuss the test M-5-brane in M-brane
backgrounds.
There are two new features which do not emerge in the previous
cases of the test M-2-brane: one is the fact that the M-5-brane
action contains worldvolume self-dual 2-form gauge potential 
${\cal A}_{2}$ in addition to the usual scalar 
fields\cite{pst1}\cite{sch1}.
The super-transformation of ${\cal A}_{2}$ is determined by
the requirement of the invariance of the ``modified'' field strength
\cite{tow5} 
given by 
\beqa
{\cal H}=d{\cal A}_{2}-C^{(3)}.
\eeqa
The other is the introduction of the superspace 6-form field strength
$C^{(6)}$ \cite{leh1} whose field strength takes the form 
\beqa
R^{(7)} & \equiv & dC^{(6)}-\frac{1}{2}C^{(3)}R^{(4)}\nonumber \\
& = & \frac{i}{5!}
E^{\hat{m_{1}}}...E^{\hat{m_{5}}}
\bar{E^{\hat{\alpha}}}
(\Gamma_{\hat{m_{5}}...\hat{m_{1}}})
^{\hat{\alpha}\hat{\beta}}E^{\hat{\beta}}
+\frac{1}{7!}E^{\hat{m_{1}}}...E^{\hat{m_{7}}}
F_{\hat{m_{7}}...\hat{m_{1}}}^{(7)}
\eeqa
where the 7-form $F^{(7)}$ is the Hodge dual of the bosonic 4-form
field strength.

(3a)the M-5-brane in the M-2-brane background

First, we will consider the test M-5-brane in the M-2-brane
background (\ref{M2back}).
This set-up might seem to be unreasonable
because M-5-branes cannot have 
the boundary\cite{tow11}.
However, test branes have no boundary in this method as stated
above in (2b) case.
Thus, we can compute the
superalgebra and discuss supersymmetric configurations in this case
in the same way.
The M-5-brane action is\cite{pst1}
\beqa
S_{M5} &=& -T\int d^{6}\xi [ 
\sqrt{-{\rm det}(g_{ij}+\tilde{{\cal H}}_{ij})}
+\frac{\sqrt{-g}}{4(\partial a)^{2}}(\partial_{i} a)
({\cal H})^{\ast ijk}{\cal H}_{jkl}(\partial^{l} a) ]
\nonumber\\
& &+\int (C^{(6)}+\frac{1}{2}{\cal H}C^{(3)})
\eeqa
where $({\cal H})^{\ast ijk}=\frac{1}{3!\sqrt{-g}}
\varepsilon^{ijki'j'k'}{\cal H}_{i'j'k'},\ 
\tilde{{\cal H}}^{ij}=\frac{1}{\sqrt{-(\partial a)^2}}
({\cal H})^{\ast ijk}\partial_{k}a$ and $a$ is an auxiliary worldvolume 
scalar field. 
Since the superspace 1-form and the (transformation of )
$C^{(3)}$ in the M-2-brane
background are already given by
(\ref{E1M2}) and (\ref{C3M2}),
the transformation of ${\cal A}_{2}$ is
\beqa
\delta {\cal A}_{2}=-\frac{i}{2}H^{1/3}dy^{c}\delta_{c}^{\hat{a}}
 dy^{d}\delta_{d}^{\hat{b}}\bar{\varepsilon}^{+}\Gamma_{\hat{a}\hat{b}}
\theta^{+}(\equiv \bar{\varepsilon}^{+} \Delta_{2} )
\eeqa
and the transformation of $C^{(6)}$ is deduced in the same way
as the 3-form $C^{(3)}$. 
As a result, we get $\delta {\cal L}^{WZ}\equiv 
d(\bar{\varepsilon}^{+} \Delta) $ where\footnote{
Because of the invariance of $R^{(7)}$ under the super-transformation,
$\delta C^{(6)}-\frac{1}{2}C^{(3)}\delta C^{(3)}$ in the
background can be written as
a d-exact form $d(\bar{\varepsilon}^{+}\Delta_{5})$.
Then, it is shown that it holds $\delta {\cal L}^{WZ}
=d(\bar{\varepsilon}^{+}\Delta_{5}-\frac{1}{2}d{\cal A}_{2}
\delta {\cal A}_{2})$ to {\it full} order in $\theta$.}
\beqa
\Delta^{\alpha}=-\frac{i}{4!}H^{1/3}dx^{\nu}\delta_{\nu}^{\mu} 
dy^{b_{1}}\delta_{b_{1}}^{\hat{a_{1}}}...
dy^{b_{4}}\delta_{b_{4}}^{\hat{a_{4}}}
(\Gamma_{\hat{a_{4}}...\hat{a_{1}}\mu}\theta^{+})^{\alpha}
-\frac{i}{4}H^{1/3}dy^{b_{1}}\delta_{b_{1}}^{\hat{a_{1}}}
dy^{b_{2}}\delta_{b_{2}}^{\hat{a_{2}}}d{\cal A}_{2}
(\Gamma_{\hat{a_{2}}\hat{a_{1}}}\theta^{+})^{\alpha}.
\eeqa
The supercharge is given as before by
$Q_{\alpha}^{+} \equiv Q_{\alpha}^{+(0)}-i\int_{{\cal M}_{5}}
({\cal C}\Delta)_{\alpha} $
where $Q_{\alpha}^{+(0)} $ takes the form\cite{tow5} 
\beqa
Q_{\alpha}^{+(0)} =\int_{{\cal M}_{5}} d^{5}\xi [ i\Pi_{\alpha}^{+}
-\Pi_{\mu}({\cal C}\Gamma^{\mu}\theta^{+})_{\alpha}
+\frac{i}{2}{\cal P}^{\underline{i}\underline{j}}
({\cal C}(\Delta_{2})_{\underline{i}\underline{j}})_{\alpha} ]
\eeqa
where $\underline{i}$ is the space index of the test M-5-brane
worldvolume and $\Pi_{\mu}, \Pi_{\alpha}^{+}$ and 
${\cal P}^{\underline{i}\underline{j}}$ are the conjugate momemta of 
$x^{\mu}$, $\theta^{+}$ and 
${\cal A}_{\underline{i}\underline{j}}$, respectively.

Thus, the superalgebra is obtained as
\beqa
\{ Q_{\alpha}^{+}, Q_{\beta}^{ +}\}= 2\int_{{\cal M}_{5}} d^{5}\xi
\Pi_{\mu}
({\cal C}\Gamma^{\mu})_{\alpha\beta} 
+2\int_{{\cal M}_{5}} d^{5}\xi
\frac{i}{2}{\cal P}^{\underline{i}\underline{j}}
H^{1/3}\partial_{\underline{i}}y^{c}\delta_{c}^{\hat{a}}
\partial_{\underline{j}}y^{d}\delta_{d}^{\hat{b}}
({\cal C}\Gamma_{\hat{a}\hat{b}})_{\alpha\beta}\nonumber \\
-2 \cdot\frac{1}{4!}T\int_{{\cal M}_{5}} 
dx^{\mu} dy^{a_{1}}...dy^{a_{4}}
({\cal C}\Gamma_{\mu a_{1}...a_{4}})_{\alpha\beta}
-2\cdot\frac{1}{4}T\int_{{\cal M}_{5}}
dy^{a}dy^{b}d{\cal A}_{2}
({\cal C}\Gamma_{ab})_{\alpha\beta}
+{\cal O}(\theta^{2})\label{m5m2salg}
\eeqa    
The third term in (\ref{m5m2salg}) means the string intersecion
with the M-2-brane background leads to the preservation of
1/4 supersymmetry, which is again consistent
with ref.\cite{tow9}, ref.\cite{tey1}
and the result of (2b) case. 
And if we choose the temporal gauge $a(\xi)=t $,
the second term including ${\cal P}_{\underline{i}\underline{j}}$
turns to be 
the same as the last term as in ref.\cite{tow5}.
Thus, these terms imply the possibility of
the triple intersection with the configuration 
\begin{center}
\begin{tabular}{lllllcrrrrr}
background M-2 & 1 & 2 & - & - & - & - & - & - & - & - \\
test M-5 & - & 2 & 3 & 4 & 5 & 6 & - & - & - & - \\
M-2 & - & - & 3 & 4 & - & - & - & - & - & - 
\end{tabular}
\end{center}
where the third M-2 brane is within the M-5-brane  
(to form the bound state)\cite{tow5}.
This configuration would preserve 1/4
spacetime supersymmetry since 
${\cal C}\Gamma_{\hat{2}\hat{3}\hat{4}\hat{5}\hat{6}} $ anticommutes
with ${\cal C}\Gamma_{\hat{3}\hat{4}} $.
We note that the possibility of this triple intersection is discussed
directly on the basis of 
the spacetime superalgebra, by making use of the fact that
{\it there exist an M-brane} as a background {\it from the beginning}.  

Next, we will consider the M-2-brane/M-5-brane bound state.
Suppose $\xi^{1}=x^{1},\xi^{2}=x^{2}$
(that is, a two-dimensional worldvolume subspace of the test
M-5-brane is parallel to the background M-2-brane).
Then, if $ {\cal A}_{2}=0$,
all the terms except 
$\Pi_{0}{\cal C}\Gamma^{0}(\propto H^{-1/6})$
in the r.h.s. of (\ref{m5m2salg})
vanish in the static gauge. 
However, since the harmonic function depends on the
``relatively transverse'' coordinates
$\xi^{3},\xi^{4}$ and $\xi^{5},\ 
\int d^{5}\xi H^{K}$ do not go to zero even in the limit of
vanishing distance.
This means that
we cannot deduce the M-2-brane/M-5-brane bound state from the
algebra at least straightforward.
But if we consider the background whose harmonic function
does not depend on the ``relatively transverse'' coordinates
as in ref.\cite{tow9}, the r.h.s. of the algebra vanishes
in the limit, and we can
obtain the 1/2-supersymmetric M-2-brane/M-5-brane bound state.

(3b)the M-5 brane in the M-5-brane background

Finally we will deal with the 
test M-5 brane in the M-5-brane background(\ref{m5back}).
Since almost all the preparations have been already given above
and the precedures are similar to before, 
we present only the transformation of ${\cal A}_{2}$,
the result of the superalgebra as well as
its interpretations in this case.
The transformation of ${\cal A}_{2}$ in the M-5 brane background is 
\beqa
\delta {\cal A}_{2}=-iH^{1/6}dx^{\nu}\delta_{\nu}^{\hat{\mu}}
dy^{b}\delta_{b}^{\hat{a}}\bar{\varepsilon}^{+}\Gamma_{\hat{\mu}\hat{a}}
\theta^{+}+{\cal O}(\theta^{3}).
\eeqa 
The superalgebra is give by 
\beqa
\{ Q_{\alpha}^{+}, Q_{\beta}^{+}\}= 2\int_{{\cal M}_{5}} d^{5}\xi
\Pi_{\mu}
({\cal C}\Gamma^{\mu})_{\alpha\beta} 
+\frac{2}{2}\int_{{\cal M}_{5}} d^{5}\xi
i{\cal P}^{\underline{i}\underline{j}}
H^{1/6}\partial_{\underline{i}}x^{\nu}\delta_{\nu}^{\hat{\mu}}
\partial_{\underline{j}}y^{b}\delta_{b}^{\hat{a}}
({\cal C}\Gamma_{\hat{\mu}\hat{a}})_{\alpha\beta}\nonumber \\
-\frac{2}{12}T\int_{{\cal M}_{5}} 
dx^{\mu_{1}}...dx^{\mu_{3}} dy^{a_{1}}dy^{a_{2}}
({\cal C}\Gamma_{\mu_{1}...\mu_{3} a_{1}a_{2}})_{\alpha\beta}
-\frac{2}{4!}T\int_{{\cal M}_{5}} 
dx^{\mu}dy^{a_{1}}...dy^{a_{4}}
({\cal C}\Gamma_{\mu a_{1}...a_{4}})_{\alpha\beta}\nonumber \\
-\frac{2}{2}T\int_{{\cal M}_{5}}
dy^{a}dx^{\mu}d{\cal A}_{2}
({\cal C}\Gamma_{\mu a})_{\alpha\beta}
+{\cal O}(\theta^{2}).
\eeqa
In fact only the fourth term is difficult to derive straightforward in 
this case
because the magnetic 3-form potential
which cannot be globally expressed 
contributes to $\delta {\cal L^{WZ}} $.
However, since they can be expressed locally in a certain gauge,
we can confirme in a gauge that some of these terms do not vanish,
although we do not exhibit the computation 
because is very
primitive and awkward to present.
Then, we reproduce the term on the ground that it should be 
gauge invariant
and Lorentz (SO(1,5)$\times$SO(5)) covariant.

Now, we will interpret the result. 
If the test M-5-brane is oriented paralell to the background M-5-brane,
perfectly the same circumstances occur as in the case of the M-2-brane 
paralell to the background M-2-brane.
So, we present only the result,
without repeating the explanations.
The configuration of two parallel M-5-branes and
an M-5-brane/M-5-brane bound state, both of which preserve
1/2 spacetime supersymmetry, are deduced.
If the test 5-brane intersects on 3-brane 
with the background M-5-brane,
1/4 spacetime supersymmetry is preserved
because of the third term (as in ref.\cite{tow9}).
The fourth term shows that
the string intersection of the two M-5-brane  
leads to the preservation of 1/4 supersymmetry\cite{izq1}. 
The last term, together with the second term with the temporal gauge
condition $a(\xi)=t $,  
implies the possibility of the existence of
the triple intersections preserving 1/4 supersymmetry 
with the following configurations: 
\begin{center}
\begin{tabular}{lllllcrrrrr}
background M-5 & 1 & 2 & 3 & 4 & 5 & - & - & - & - & - \\
test M-5 & 1 & 2 & 3 & - & - & 6 & 7 & - & - & - \\
M-2 & - & - & 3 & - & - & 6 & - & - & - & - 
\end{tabular}
\end{center}
and
\begin{center}
\begin{tabular}{lllllcrrrrr}
background M-5 & 1 & 2 & 3 & 4 & 5 & - & - & - & - & - \\
test M-5 & - & - & - & - & 5 & 6 & 7 & 8 & 9 & - \\
M-2 & - & - & - & - & 5 & 6 & - & - & - & -
\end{tabular}
\end{center}
where each of the third M-2 branes is within the test M-5-brane. 

%
\section{Discussion}
In summary we have discussed the method of computing explicitly 
the spacetime superalgebras
from the test M-brane actions in {\it M-brane backgrounds}
to the lowest order in $\theta$.
As the conseqences we have {\it derived} all the 
1/4-supersymmetric intersections of the two M-branes known before
from the central charges
of the {\it spacetime} superalgebras,
as the supersymmetric ``gauge fixing'' of the test brane 
permitted in the background.
We have also deduced some of 1/2 supersymmetric bound states
of two M-branes from examining
the behavior of the harmonic functions in the limit of vanishing
distances.
In addition, the possibilities of
some triple intersections preserving 1/4 supersymmetry have
been discussed.

In order to obtain the 1/2 supersymmetric bound state
(M-2-brane within M-5-brane),
we need to assume 
{\it only} in the case of
the M-5-brane in the M-2-brane background that the hamonic function is
independent of the ``relatively transverse coordinates.
Thus, when we deal with the system composed of 
a p-brane and a q-brane with the inequality
$q \ge p$, it is more suitable for this method to discuss the system as
the p-brane in the q-brane background.
Except for this subtlety, the method discussed
in this paper is confirmed to be consitent with
those presented before, and hence reliable.

Having confirmed its reliability,
we will apply this method to
the p-branes in 10-dimensional {\it massive} IIA backgrounds\cite{sato1}
as stated in the introduction,
in which case {\it the background have to be nontrivial}
\cite{rom1}\cite{berg3}.
It will also be interesting to apply this to the cases of
other backgrounds\cite{sato2}, or
use it to the new supersymmetic invariant
p-brane action
presented recently\cite{berg1}.

\parbigskipn

{\Large\bf Acknowledgement}

\parbigskipn
I would like to thank Prof. J. Arafune for careful reading of the
manuscript and useful comments. I am grateful to Akira Matsuda for
many useful discussions and encouragement. 
I also would like to thank Taro Tani for stimulating discussions.

\parbigskipn\parsmallskipn
%
%

\end {document}